# Electronic and vibrational spectroscopy of miscible MgO-ZnO ternary alloys

K. Aziz[1,2] and C.E. Ekuma[1, *]

[1]*Department of Physics, Lehigh University, Bethlehem, PA 18015, USA*
[2]*Department of Physics, Wesleyan University, Middletown, 06459 CT, USA*
(Dated: January 30, 2020)

The ordered structure of MgO-ZnO alloy system is a versatile tunable optical material promising for diverse optoelectronic applications. However, isovalent and isostructural alloy compositions of MgO-ZnO are generally unstable at ambient conditions. Using state-of-the-art *ab initio* evolutionary simulations, we predict and study the properties of stable phases of MgO-ZnO. We establish the dynamical stability of the predicted crystal structures through the phonon and Raman spectroscopy. Detailed analyses of two of the most stable structures reveal highly tunable properties that could be explored for photonic and optical applications.

## I. INTRODUCTION

Zinc oxide (ZnO) is a wide, direct bandgap $E_g \sim 3.4$ eV II-VI semiconductor that has attracted increased interest for both fundamental research and diverse optoelectronic applications.[1–3] For example, ZnO is a promising material for light-emitting and laser diodes operating in blue and ultraviolet (UV) regime.[2–4] However, low electroluminescence often diminish the efficiency of ZnO-based devices.[5,6] To realize and enhance the efficiency of ZnO-based devices, the electronic structure through band-engineering can be systematically tuned by doping. ZnO can be alloyed with MgO to form a ternary $Mg_xZn_{1-x}O$ compound thus, enabling bandgap engineering and luminescence in the UV regime.[7] Often, the combination of these group-II oxides in alloys leads to crystal structure mismatch: undoped ZnO prefers the hexagonal wurtzite (B4) structure or the four-fold coordinated zincblende structure (B3) while MgO favors the cubic rocksalt (B1) structure at ambient conditions.[8,9] Experiments have shown that $Mg_xZn_{1-x}O$ exhibits the B4 structure for high ZnO concentration[10] while preferring the B1 structure at high MgO concentration.[11] At intermediate concentrations, it exhibits phase separation[12] due to compositional gradient, which often leads to thermodynamically unstable crystal structure. In general, the isovalent and isostructural II-VI alloys are thermodynamically unstable because the mixing enthalpy in either the B1, B3 or B4 structure is always positive.[12,14]

The difficulty in synthesizing stable phases of MgO-ZnO alloys has hindered exploiting adequately their promising properties for technological applications. This prompted earlier calculations on the stability of the $Mg_xZn_{1-x}O$ alloy system[12,13]. For example, the work of Sanati *et al*[12] on various alloys of MgO-ZnO found energetic stability, under certain conditions, in the sixfold-coordinated structure for Zn concentrations below 67%. Instead of the traditional doping approach of alloying ZnO with MgO to obtain often phase immiscible $Mg_xZn_{1-x}O$ alloy, we can apply high-throughput crystal structure prediction techniques to search for energetically and dynamically stable stoichiometric MgO-ZnO alloy systems. A high-throughput experiment to achieve this is, however, daunting because of the ample parameter space that needs to be explored. Computationally, such an approach is now routinely used to predict new crystal structures,[15,16] including recently, the prediction of energetically stable high pressure phase of $Mg_xZn_{1-x}O$.[17] Herein, we report the electronic structure and thermodynamic stability of the stable phases of $MgZnO_2$ crystal structure at ambient conditions obtained from the recently developed evolutionary algorithm for the prediction of crystal structure.[16] We provide detailed physical properties and characterize the electronic, mechanical, phononic, and optical spectroscopy. We extensively discuss the results in relation to experimental data of the parent systems and provide details of the crystal structure information (see Supplementary Material). We hope that they will motivate future experimental investigations of the electronic structure, particularly using optical measurements and photoemission, and the vibrational spectroscopy by Raman measurements.

## II. METHOD

The ground state crystal structures of $MgZnO_2$ are predicted using the *ab initio* evolutionary algorithm CALYPSO.[16] We performed variable compositional simulations for $Mg_nZn_mO_{m+n}$ systems (n, m $\leq$ 30) while maintaining the stoichiometric composition of $MgZnO_2$. The structural optimizations and electronic structure calculations for the selected structures are performed with the *Vienna Ab Initio Simulation Package* (VASP) code.[18] The electronic wavefunctions of the systems are represented by a planewave basis with an energy cutoff of 550 eV. The exchange-correlation interactions were described with the HSE06 hybrid functional[19] with a $10 \times 10 \times 10$ $\Gamma-$centered $\mathbf{k}-$point grid, which ensures the energy (charge) is converged to within $10^{-4}$ ($10^{-8}$) eV. We also obtained the electronic structure using the strongly constrained and appropriately normed (SCAN) semilocal density functional.[20] (see the Supplementary Material). The phononic and vibrational spectroscopy is obtained using density functional perturbation theory (DFPT)[21] as implemented in VASP in conjunction with

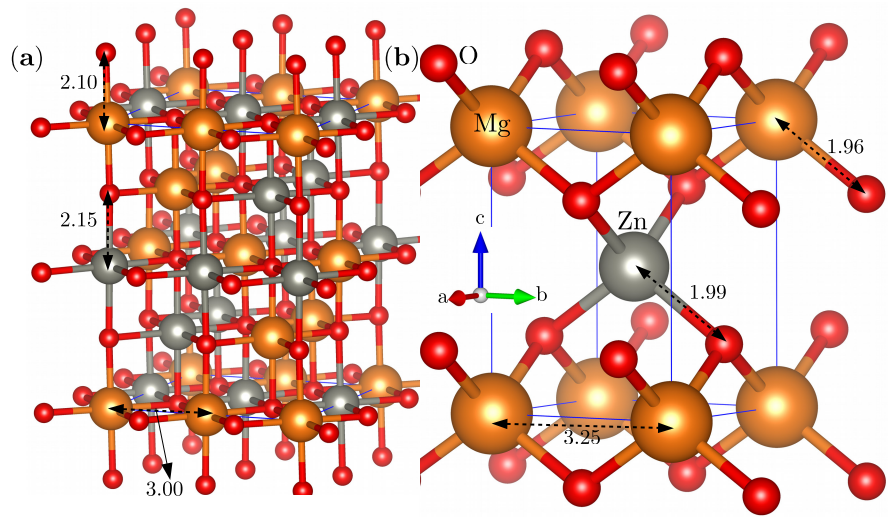

FIG. 1. **Crystal structure of MgZnO$_2$.** The crystal lattice of structure **A** (a) and structure **B** (b).

the Phonopy package.[22]

### III. RESULTS AND DISCUSSION

*Crystal structure and stability–* In this work, we present two stable ground state crystal structures of MgZnO$_2$. The two of the predicted structures both belong to the tetragonal crystal symmetry $I4_1/amd$ space group No. 141 and $P\bar{4}m2$ space group No. 115, respectively (see, Figure 1). We designate the former and latter structures as **A** and **B**, respectively. The predicted lattice constants, along with other salient parameters, are presented in Table I and the Wyckoff positions in the crystallographic information file format are included in the Supplementary material. The phase stability of the MgO-ZnO systems are studied by calculating the formation energy $E_{\text{form}}$ per atom; $E_{\text{form}}$ is a measure of the energetic stability of a material and for our system, it is obtained as $E_{\text{form}} = E_{\text{tot}} - \sum_i c_i \mu_i$, where $i$ denotes different types of atoms present in the unit cell of the material with concentration $c_i$, and $\mu_i$ is the chemical potential approximated with the standard state (bulk) energy $E_i$ of the corresponding atom, $i$. The two structures exhibit negative formation energies with an average $E_{\text{form}} \sim -2.11\,\text{eV/atom}$; this implies an exothermic process confirming the phase and energetic stability of the crystal. To further confirm the energetic stability for the predicted structures, we calculate the enthalpy profile as $\Delta\text{H} = \Delta\text{G} + \text{T}\Delta\text{S}$, where H, G, S, and T is enthalpy, free energy, entropy, and temperature, respectively, for DFT, T= 0 K) as a function of generations (see Figure S1 in the Supplementary Material). All the predicted structures show a negative formation enthalpy $\Delta H >$-5.0 eV, which supports the energetic stability of the structures. The predicted $\Delta H$ is consistent with previous theoretical results for the Mg$_{1-x}$Zn$_x$O system[12,13]; though these previously reported $\Delta H$ are negative, they are smaller than our results due to different stoichiometric compositions.

While the energetic stability is often a confirmation that a crystal can be experimentally synthesized, in most cases, basing the overall stability on the energy alone may be misleading. In the absence of any phase transition, a crystal may have negative formation energy yet thermodynamically unstable. To ascertain the dynamical stability of the predicted structures, we performed phonon and vibrational spectroscopy calculations (Figure 2). We also show in Figure 2 the Raman and Infrared spectroscopy. Both structures show positive definite phonon frequency along the various high symmetry point in the first Brillouin zone, confirming the dynamical stability. From the phonon atomic projected density of states, the acoustic branch is dominated by Zn atoms, while the optical branch is composed of a strong hybridization between the O and Mg atoms. This trend is consistent with the domination of the frequency scale of the acoustic (optical) phonon modes by atoms with larger (smaller) masses similar to a diatomic linear chain model.[23]

The predicted crystal structures belong to the tetragonal crystal symmetry, albeit with different point group symmetry leading to a different irreducible representation of the vibrational modes. Crystal structure **A** belongs to the point group D$_{4h}$ (4/mmm) with sixteen atoms in the primitive cell. The irreducible representation of the corresponding forty eight normal vibrational modes at the Brillouin zone center (Γ-point) is $\Gamma = 2A_{1g}(R) + 3A_{1u} + A_{2g} + 6A_{2u}(IR) + 2B_{1g}(R) + 3B_{1u} + B_{2g}(R) + 6B_{2u} + 9E_u(IR) + 3E_g(R)$, where IR and R is the infrared and Raman active modes. A$_{1u}$, A$_{2g}$, B$_{1u}$, and B$_{2u}$ are the silent mode. The R-active E$_g$, B$_{1g}$, and the A$_{1g}$ modes are mainly due to the symmetric stretching vibration, the symmetric bending vibration, and the antisymmetric bending vibration of the O-

TABLE I. Some important predicted parameters of MgZnO$_2$. [a]

| Name | a/c | SG | $E_g$ | $m_e(\parallel / \perp)$ | $m_h(\parallel / \perp)$ | B | E | G | M | $\mu$ | $\zeta$ | $\Theta_D$ | $\epsilon_\infty(\parallel / \perp)$ | $\epsilon_0(\parallel / \perp)$ |
|---|---|---|---|---|---|---|---|---|---|---|---|---|---|---|
| A | 6.008/8.491 | 141 | 4.62(I) | 0.296/0.293 | 2.973/2.361 | 180.82 | 223.63 | 86.45 | 296.09 | 0.29 | 2.09 | 377.22 | 2.87/2.87 | 7.91/7.78 |
| B | 3.245/4.509 | 115 | 3.70(D) | 0.304/0.312 | 1.772/2.728 | 140.13 | 116.26 | 42.80 | 197.20 | 0.36 | 3.27 | 502.46 | 2.83/2.83 | 6.80/5.40 |

[a] The lattice constant includes $a$ (Å), space group SG with crystal symmetry $I4_1/amd$ ($P\bar{4}m2$) for structure **A** (**B**); the electronic properties include the energy bandgap $E_g$ (eV), electron $m_e$ and hole $M_h$ effective masses in units of free electron mass $m_0$, where $\parallel$ ($\perp$) denotes parallel (perpendicular) direction and I (D) is indirect (direct) energy bandgap; the mechanical properties include the bulk modulus B (GPa), Young's modulus E (GPa), shear modulus G (GPa), P-wave modulus M (GPa), Poisson ratio $\mu$, and bulk/shear ratio $\zeta$; the dielectric properties include the high photon-energy $\epsilon_\infty$ and static $\epsilon_0$ dielectric constants; and the Debye temperature $\Theta_D$ (K) representing the thermal property.

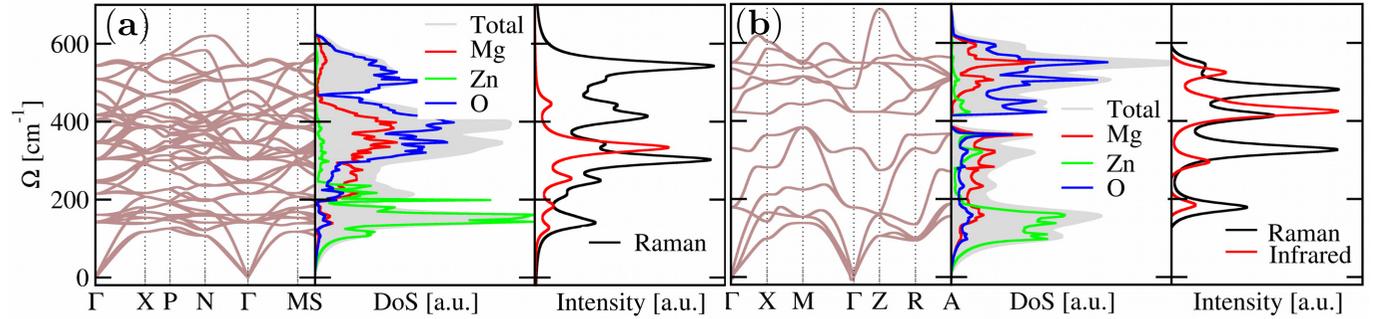

FIG. 2. **Phonon and vibrational spectroscopy of MgZnO$_2$.** Phonon spectra and the corresponding vibrational density of states, the Raman, and Infrared spectra for structure **A** (a) and structure **B** (b).

Mg/Zn-O, respectively. The phonon spectra, the phonon total and the atomic projected density of states along with the Raman and Infrared spectra are presented in Figure 2(a). (Further analysis of the irreducible representation of the phonon modes are shown in Table S1 of the Supplementary Material.) We hope the presented phonon, vibrational, Raman, and Infrared spectroscopy will guide experimental and other analyses. For the crystal structure **B** with point group symmetry $D_{2d}(-42m)$, the primitive cell contains four atoms; as such, there will be a total of twelve normal vibrational modes at the $\Gamma$ point [Figure 2(b)]. The irreducible representation of the vibrational modes at the $\Gamma$ point could be expressed as $\Gamma = A_1(R) + 3B_2(IR+R) + 4E(IR+R)$. One low-energy $B_2$ and one double $E$ modes are the usual acoustic modes characterized by the transverse acoustic (TA), longitudinal acoustic (LA), and the out-of-plane transverse acoustic or flexural mode (ZA); the remaining belong to the optical mode. Using group theory analysis[24] in conjunction with the Raman and Infrared spectra [Figure 2(b)], we assign the feature at $\sim$178.70 cm$^{-1}$ to the degenerate IR- and R-active $E$ mode and the structure at $\sim$326.28 cm$^{-1}$ to the IR- and R-active $B_2$ mode. We predict a degenerate IR- and R-active $E$ mode at $\sim$413.94 cm$^{-1}$, followed by IR- and R-active $E$ mode at $\sim$480.60 cm$^{-1}$. We predict an R-active $A_1$ mode though with small activity around 531.00 cm$^{-1}$ before a degenerate IR- and R-active $E$ mode at $\sim$548.81 cm$^{-1}$. (Further details of the irreducible representation are presented in Table S2 in the Supplementary Material.) We note that the predicted features in both the vibrational and Raman spectra depict more of ZnO character than MgO. For example, room temperature Raman spectroscopy experiments reported the R-active modes $E \sim 583$ cm$^{-1}$ and $A_1 \sim 574$ cm$^{-1}$.[25]

The Debye temperature $\Theta_D$ is an important thermal property of a material. It quantifies the temperature below which modes start to freeze out and above which, the modes gain enough energy to be excited.[26] The Debye temperature can be obtained as $\Theta_D^3 = \frac{1}{3}\sum_i \frac{1}{\Theta_{\Gamma,i}^3}$, where $\Theta_{\Gamma,i} = \hbar\Omega_{\Gamma,i}/\kappa_B$ and $\Theta_{\Gamma,i}$ corresponds to the frequency at the zone boundary of the $i$th acoustic mode.[27] The predicted $\Theta_D$ is presented in Table I, which seem to be closer to that of ZnO with $\Theta_D \sim 440$ K than to MgO with $\Theta_D \sim 948$ K as reported by experiments.[9,25]

*Elastic and mechanical properties–* The elastic properties not only provide the information on the mechanical stability of material but also serve as a guide on the potential device applications of a given material. In this regard, we have calculated the elastic properties of the studied materials using the finite difference method. We used a planewave energy cutoff of 700 eV to ensure the adequate convergence of the stress tensor. The elastic tensor matrix is determined by performing six finite distortions of the lattice and obtaining the elastic constants from the strain-stress relationship.[28,29] The predicted elastic tensor matrices are presented in Table II. Using the calculated elastic tensor matrix, we predict the elastic constants presented in Table I. Aside from the isotropic elastic behavior (the calculated elastic anisotropy ratio $A = 2C_{44}/(C_{11} - C_{12}) \sim 1.0$), the two studied structures show markedly different elastic behavior. Structure **A** (**B**) could withstand linear compressibility of ∼1.81 (1.65 TPa$^{-1}$). The predicted Bulk modulus of structure **A** is ∼ 41 GPa larger than that of structure **B**. The same trend is observed for Young's modulus E, shear modulus G, P-wave modulus M, and the Poisson ratio $\mu$, which is larger by E∼ 107.37 GPa, G∼ 43.65 GPa, M∼ 98.89 GPa, and $\mu \sim 0.07$. Though structure **A** exhibits large elastic constants, structure **B** is, however, more ductile from the bulk/shear ratio value (Table I).

TABLE II. Calculated elastic constant tensor $C_{ij}$ for MgZnO$_2$. Only half side is shown by symmetry.

| | Structure **A** | | | | | | Structure **B** | | | | | |
|---|---|---|---|---|---|---|---|---|---|---|---|---|
| $C_{ij}$ | 1 | 2 | 3 | 4 | 5 | 6 | 1 | 2 | 3 | 4 | 5 | 6 |
| 1 | 312.05 | 88.39 | 142.04 | 0.36 | 0.0 | 0.0 | 225.77 | 85.46 | 123.68 | 0.0 | 2.32 | 0.0 |
| 2 | | 313.29 | 143.02 | 0.38 | 0.0 | 0.0 | | 227.90 | 124.45 | 0.0 | 2.41 | 0.0 |
| 3 | | | 255.17 | 0.45 | 0.0 | 0.0 | | | 151.77 | 0.0 | 2.40 | 0.0 |
| 4 | | | | 114.33 | 0.0 | 0.0 | | | | 70.48 | 0.0 | 0.06 |
| 5 | | | | | 113.96 | 0.0 | | | | | 70.61 | 0.0 |
| 6 | | | | | | 58.38 | | | | | | 20.55 |

*Electronic and transport properties–* The description of the electronic properties of the two forms of MgZnO$_2$ studied herein as obtained using the HSE06 hybrid functional is provided by the band structure, total, and atomic projected density of states in Figure 3. We predict an indirect energy bandgap $E_g \sim 4.62$ eV along the $\Gamma - N$ point of the high symmetry point of the Brillouin zone for structure **A**. We note that the direct energy bandgap is 4.52 eV, which is only 0.10 eV larger than the smallest bandgap; as such, the indirect and direct energy bandgaps could be said to be almost degenerate. Structure **B** is predicted to be a direct bandgap semiconductor with $E_g \sim 3.70$ eV at the $\Gamma-$point. The predicated bandgaps are listed in Table I. We note that the energy bandgap predicted with the semilocal SCAN (see Figure S2 in the Supplementary material) and the Perdew-Burke-Ernzerhof generalized gradient approximation (PBE-GGA)[30] is significantly smaller in both materials. For example, for structure **A**, PBE-GGA and SCAN functional led to $E_g \sim 3.32$ and 2.73 eV, respectively. We obtained $E_g \sim 2.43$ and 1.98 eV using SCAN and PBE-GGA functional, respectively, for structure **B**.

The two studied structures share in common some of the features of ZnO. For example, the predicted energy bandgap is in the same range as the ∼3.4 eV reported for bulk ZnO by several experiments and computations.[9,25,31–33] Aside from the difference in the energy bandgap of the two structures, the character of the states in the proximity of the Fermi energy are similar for the two structures. From the projected density of states, the lower-lying group in the valence states consists of highly hybridized O (from s- and p-states) while the upper group is mostly dominated by Zn (d-states), O (p-states), and tiny contribution from Mg (s- and p-states). The upper-most conduction states are dominated by a strong hybridization between Zn (s-states), O (s- and p-states). This is evident from the free electron-like (s-like) states at the conduction band minimum (CBM) located at the $\Gamma-$point. The total width of the upper valence bands within 10.0 eV from the valence band maximum (VBM) is 6.30 (5.28 eV) for structure **A** (**B**). This value is in good agreement with ∼ 6.50 eV for MgO and approximately 1-2 eV smaller than the ∼ 7.2 eV for ZnO reported by experiments.[31,34] The total width of the entire valence states is 19.52 (18.25 eV) for structure **A** (**B**). These values are within the experimental uncertainty of the total valence width of ZnO.[25] We note that the valence bandwidth obtained with the scan functional decreased by $\sim 5.2\%$. With the PBE-GGA functional, the decrease is more significant $\sim 14.07\%$.

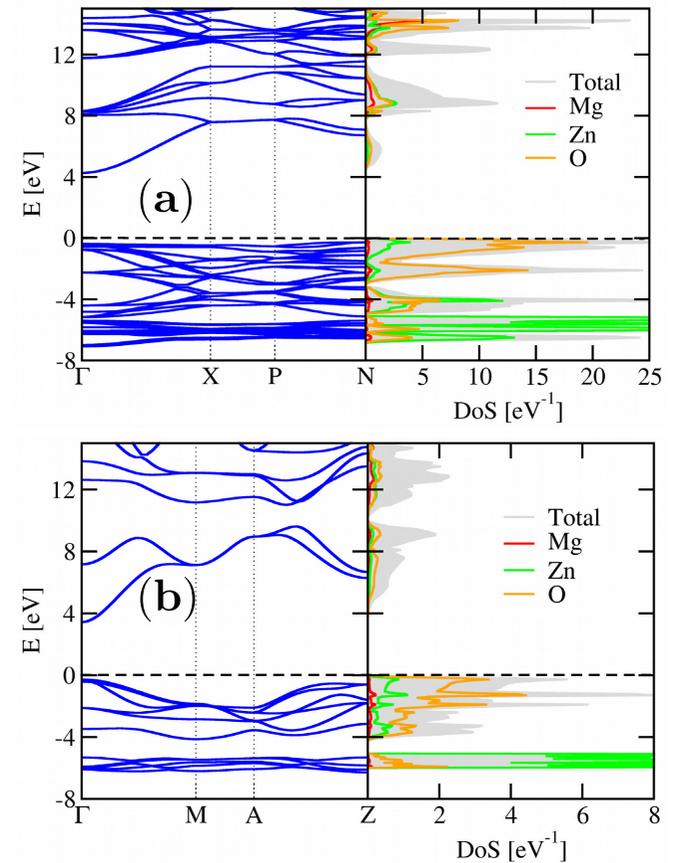

FIG. 3. **Electronic properties of MgZnO$_2$**. Electronic band structure and the corresponding electron density of states for structure **A** (a) and structure **B** (b). The horizontal dashed line is the Fermi energy, which is set at zero of the energy scale.

The effective mass is a measure of the curvature of the calculated band structure, and it provides insight into the transport properties of materials. To this end, we calculate the carrier band effective mass of the two studied structures. The band effective mass $m^b$ is obtained from the band structure by fitting a parabola $E_k = \frac{\hbar^2}{2m_0}\vec{k}^T A \vec{k}$ to the states around the band extremum (CBM and VBM) of Figure 3, where $k = (\parallel, \perp)$ is the parallel and perpendicular direction of the $k$-point measured from the band extremum, and $m_0$ is the free electron mass. The predicted electron and hole effective masses along the $\parallel$ ($\perp$) direction are listed in Table I. The effective mass decreased by $\sim 6.50$ and $15.20\%$, respectively, with the SCAN and PBE-GGA functionals. The large hole effective mass, especially for structure **A**, is due to the rather flat bands around the top of the valence states. We note that the predicted $m_e^{\parallel/\perp}$ straddle that of ZnO $\sim 0.275$ and MgO $\sim 0.370$ $m_0$.[9,25,35,36] While the predicted $m_h^{\parallel}$ is in good agreement with $\sim 0.60 m_0$ reported for ZnO,[25] the calculated $m_h^{\perp}$ is in better agreement with the 1–2 $m_0$ reported for MgO.[36]

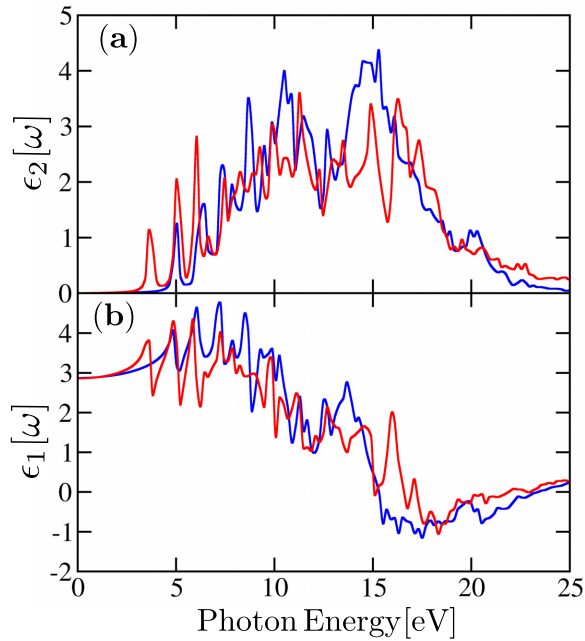

FIG. 4. **Optical properties of MgZnO$_2$**. Calculated imaginary part $\epsilon_2(\omega)$ (a) and the real part $\epsilon_1(\omega)$ (b) of the dielectric function of MgZnO$_2$. Blue (red) color denotes structure **A** (**B**).

*Optical spectroscopy*– To gain further insights into the physical properties of the predicted structures, we calculate the electrodynamical response as a function of photon energy ($\hbar\omega$). The optical spectroscopy provides a detailed and direct information on charge dynamics and "bulk" electronic structure of materials since it probes both free carriers and interband excitation. The optical response is determined using the spectra from our electronic structure calculations obtained with HSE06 hybrid functional without any shift or scissor operator. We used a tiny small broadening parameter $\sim 10^{-4}$ eV as such, our optical spectra show a lot of features. Using random phase approximation, we compute the real part of the complex dynamical dielectric function

$$\epsilon_{\alpha\beta}^{(2)} = \frac{4\pi^2 e^2}{\Omega} \lim_{q \to \infty} \frac{1}{q^2} \sum_{c,v,k} \Xi\, 2 w_k\, \delta(\varepsilon_{ck} - \varepsilon_{vk} - \omega),$$

where $\Xi = \langle u_{ck} + e_{\alpha\,q} | u_{vk}\rangle \langle u_{ck} + e_{\beta\,q} | u_{vk}\rangle^*$, $\omega$ is the photon frequency, $\alpha$ ($\beta$) is band indices, $v$ and $c$ denote valence and conduction band states, and $u_{ck}$ is the cell periodicity at the $k-$point over volume $\Omega$. The dispersive (real) part of the dynamical dielectric function $\epsilon^{(1)}(\omega)$ is then obtained using the appropriated Kramers-Kronig transformation

$$\epsilon_{\alpha\beta}^{(1)} = 1 + \frac{2}{\pi}\mathcal{P}\int_0^{\infty} \frac{\epsilon_{\alpha\beta}^{(2)}(\omega')\,\omega'\,d\omega'}{\omega'^2 - \omega^2 + i\eta},$$

where $\mathcal{P}$ is the principal value.[21] The calculated optical spectra are presented in Figure 4. The onset of absorption that is determined by the absorption edge depicts the characteristic optical bandgap. This corresponds to the lowest sharp structure around 4.53 (3.12) eV for structure **A** (**B**) due to the direct optical excitation.

The predicted optical bandgap is within the numerical uncertainty of the ones obtained from our band structure calculations (Table I). The optical spectra for both materials share almost the same characteristic features. In the $\epsilon_2(\omega)$ spectra [Figure 4(a)], we found a sharp structure around 6.50 (5.06), followed by a set of close sharp features and shoulders that transcend in both structures up to 17.5 eV, before the spectra systematically decay towards zero photon energy.

In Figure 4(b), we present the calculated dispersive parts of the dynamical dielectric function $\epsilon^{(1)}(\omega)$. The main features in both structures are sharp structures around 4.95 (3.50) eV for structure **A** (**B**), series of other features between $\sim 5.20$ eV to 15.10 eV, characterized by a steep decrease that started around 7.80 eV. Then, $\epsilon^{(1)}(\omega)$ for both structures becomes negative with a minimum at $\sim 15.10$ eV. The optical spectra obtained using the electronic structure spectra from the SCAN semilocal functional though exhibits the same trend, the critical energies are significantly different (see Figure S3 of the Supplementary Material). We attribute this to the nonlocal effects that are inherent in the HSE06 hybrid functional as evident from the sharper structures in the optical spectra. We further determined the high photon energy $\epsilon_{\infty}$ and the static photon energy $\epsilon_0$ limits of the dielectric constants. $\epsilon_\infty$ is obtained as the zero-photon energy limit of the $\epsilon^{(1)}(\omega)$ spectra and the obtained values are listed in Table I. To obtain $\epsilon_0$, we note that $\epsilon_0 = \epsilon_\infty + \epsilon_{ion}$, where $\epsilon_{ion}$ is the static photon energy ($\hbar\omega = 0$) contribution of the ionic lattice dynamics to the dielectric constant. We obtained from our phonon calculations $\epsilon_{ion}(\parallel / \perp) \sim 5.04/4.91$ (3.97/2.57) for structure **A** (**B**). Using the obtained $\epsilon_{ion}$, we predict the static photon energy of the dielectric constant summarized in Table I. The predicted optical behavior of structure **A** is generally isotropic while that of structure **B** could be said to be slightly anisotropic, mainly from the dynamics of $\epsilon_{ion}(\parallel / \perp)$. We note that the $\epsilon_\infty(\parallel / \perp)$ obtained using the eigenstates from the SCAN functional as the input in our optical calculations is 3.43/3.43 (3.34/3.31) for structure **A** (**B**), which seem to be closer to the values reported for ZnO.[25] Overall, compared to the parent materials, the predicted limits of the dielectric constants straddle those of ZnO and MgO. For example, experiments report ZnO has $\epsilon_\infty(\parallel / \perp) \sim 3.75/3.70$ and $\epsilon_0(\parallel / \perp) \sim 8.75/7.80$ while MgO exhibits $\epsilon_\infty(\parallel / \perp) \sim 2.94$ and $\epsilon_0(\parallel / \perp) \sim 9.83$.[25]

### IV. CONCLUSIONS

We report detailed *ab initio* results of the structural, phononic, electronic, transport, and optical properties of newly predicted crystal structures of MgZnO$_2$ alloy system. Through the calculations of formation energies and the phonon and vibrational spectra, including the Raman and Infrared spectroscopy, we establish the dynamical stability of the predicted structures. We demonstrate highly tunable electronic and optical properties of two of the predicted structures with optoelectronic features that make them promising for application as a photonic material. We provide detailed crystal structure information to guide further experimental studies.

### V. SUPPLEMENTARY MATERIAL

See supplemental material at [URL will be inserted by publisher] for the predicted crystal lattice parameters; the irreducible representation of the phonon active modes at the Γ-point; and the predicted electronic and optical properties obtained with the SCAN functional.

### VI. ACKNOWLEDGEMENT

Supercomputer and computational resources are provided by the Lehigh University (LU) HPC Center. K. A. acknowledges support of LU Physics REU program [PHY-1852010] and C. E. E. acknowledges LU start-up and summer research fellowship.